\documentclass[11pt]{article}
\usepackage{jheppub}
\pdfoutput=1

\usepackage{slashed}
\usepackage{array,multirow}
\usepackage{graphicx,graphics}
\usepackage{amsfonts,amsmath,amssymb, bbold,bbm}
\usepackage{color,xcolor}
\usepackage{tikz} 
\usepackage{enumitem}
\usepackage{cleveref}
\usepackage{hyperref}

\newcommand{\be}{\begin{equation}}
\newcommand{\ee}{\end{equation}}
\renewcommand\({\left(}
\renewcommand\){\right)}
\renewcommand\[{\left[}
\renewcommand\]{\right]}
\newcommand{\dd}{{\rm d}}

\newcommand{\e}{{\rm e}}

\let\vec\mathbf
\let\vecS\boldsymbol

\def\E{\mathcal{E}}
\def\L{\mathcal{L}}

\def\D{\mathcal{D}}

\def\nn{\nonumber}
\def\Tr{{\rm Tr}}

\usepackage{tikz}
\usetikzlibrary{patterns}
\usetikzlibrary{arrows,shapes}
\usetikzlibrary{trees}
\usetikzlibrary{matrix,arrows} 				
\usetikzlibrary{positioning}				
\usetikzlibrary{calc,through}				
\usetikzlibrary{decorations.pathreplacing}  
\usepackage{pgffor}							

\usetikzlibrary{decorations.pathmorphing}	
\usetikzlibrary{decorations.markings}
\tikzset{
	vector/.style={decorate, decoration={snake}, draw},
	provector/.style={decorate, decoration={snake,amplitude=2.5pt}, draw},
	antivector/.style={decorate, decoration={snake,amplitude=-2.5pt}, draw},
	fermion/.style={draw=black, postaction={decorate},
		decoration={markings,mark=at position .55 with {\arrow[draw=black]{>}}}},
	fermionbar/.style={draw=black, postaction={decorate},
		decoration={markings,mark=at position .55 with {\arrow[draw=black]{<}}}},
	fermionnoarrow/.style={draw=black},
	gluon/.style={decorate, draw=black,
		decoration={coil,amplitude=4pt, segment length=5pt}},
	scalar/.style={dashed,draw=black, postaction={decorate},
		decoration={markings,mark=at position .55 with {\arrow[draw=black]{>}}}},
	scalarbar/.style={dashed,draw=black, postaction={decorate},
		decoration={markings,mark=at position .55 with {\arrow[draw=black]{<}}}},
	scalarnoarrow/.style={dashed,draw=black},
	electron/.style={draw=black, postaction={decorate},
		decoration={markings,mark=at position .55 with {\arrow[draw=black]{>}}}},
	bigvector/.style={decorate, decoration={snake,amplitude=4pt}, draw},
}

\title{Resummation and cancellation of the VIA source in electroweak baryogenesis
}

\abstract{ We re-derive the vev-insertion approximation (VIA) source
  in electroweak baryogenesis. In contrast to the original derivation,
  we rely solely on 1-particle-irreducible
self-energy diagrams.  We solve the Green's function equations both
perturbatively and resummed over all vev-insertions.  
The VIA source corresponds to the leading order
contribution in the gradient expansion of the Kadanoff-Baym (KB)
equations.  We find that it vanishes both for bosons and fermions, 
both in the perturbative and in the resummed approach.
Interestingly, the non-existence of the source is a result of
a cancellation between different terms in the KB equations,
and not of a pathology in the
vev-insertion approximation itself.  }

\author[a]{Marieke Postma,}\author[b]{Jorinde van de Vis} \author[c]{and Graham White}

\emailAdd{mpostma@nikhef.nl, jorinde.van.de.vis@desy.de, graham.white@ipmu.jp}
\affiliation[a]{Nikhef,
Science Park 105,
1098 XG Amsterdam, The Netherlands}
\affiliation[b]{Deutsches Elektronen-Synchrotron DESY, Notkestr. 85, 22607 Hamburg, Germany}
\affiliation[c]{Kavli IPMU (WPI), UTIAS, The University of Tokyo, Kashiwa, Chiba 277-8583, Japan}

\preprint{Nikhef 2022-007, DESY-22-092, IPMU22-0034}

\begin{document}

\maketitle


\newpage
\section{Introduction}

In electroweak baryogenesis (EWB) the observed matter-antimatter asymmetry
in the Universe is created during a first order electroweak phase
transition \cite{Kuzmin:1985mm,Shaposhnikov:1987tw,McLerran:1990zh,Cohen:1990it} (for a review, see \cite{Cohen:1993nk,Trodden:1998ym,Cline:2006ts,Morrissey:2012db,White:2016nbo,Garbrecht:2018mrp}). In order to satisfy the Sakharov
conditions (violation of baryon number, charge (C) and charge-parity
(CP) symmetry, and out-of-equilibrium dynamics), new physics at the
electroweak scale is required \cite{Sakharov:1967dj}. Due to the relatively low energy scale, these new interactions can be probed by experiments. For example, the latest ACME results \cite{ACME:2018yjb}
on electric dipole moments put stringent constraints on CP
violation in effective Yukawa interactions between the Higgs field and
the Standard Model (SM) fermions, already ruling out a large class of
models (see e.g. \cite{Cline:2011mm, Liebler:2015ddv, Dorsch:2016nrg, Cline:2017jvp,deVries:2017ncy, Bodeker:2020ghk}). 

Unfortunately, the  theoretical uncertainties in the computation of the baryon
asymmetry are large, as different approaches lead to predictions that
may vary by orders of magnitude \cite{Cline:2017jvp, Cline:2020jre, Cline:2021dkf}. The main difference between methods
is in the derivation of the source terms in the transport equations.
In particular, using the vev-insertion approximation (VIA),
models with CP-violating Yukawa couplings may still be marginally
consistent with the electric dipole bounds
\cite{DeVries:2018aul,Fuchs:2020uoc}, whereas in a semi-classical
computation the corresponding asymmetry is orders of magnitude too
small \cite{Cline:2021dkf}.  In this paper we will take a careful look
at the derivation of the VIA source term in the transport equations.

The transport equations describe the evolution of the phase space
densities of the plasma particles as they interact with the expanding
bubbles during the electroweak phase transition. The Higgs vacuum expectation value (vev) varies
from zero in the false vacuum outside the bubble to some finite value
inside.\footnote{In multi-field or multi-step transitions some combination of the scalar fields in the extended Higgs sector
  has a changing vev inside the bubble wall.}  This leads to a
spacetime-dependent mass for the SM fermions and other degrees of
freedom with couplings to the Higgs field, and consequently to a
semi-classical force \cite{Joyce:1994zt,Cline:2000nw,Kainulainen:2001cn, Prokopec:2003pj, Prokopec:2004ic}. In a multi-flavor set-up the mixing matrix
between the flavor and mass eigenstates varies inside the bubble wall
as well \cite{Konstandin:2004gy, Konstandin:2005cd, Cirigliano:2011di}. In the presence of CP violation, both effects can give a
chiral asymmetry in front of the bubble wall, which is generated 
by appropriate source terms in the transport equations. Electroweak sphaleron
transitions then transform the chiral asymmetry into a baryon
asymmetry.

The vev-insertion approximation treats the space-time-dependent mass
term as an interaction term, and calculates the source term in a
perturbative expansion \cite{Riotto:1995hh,Riotto:1997vy}. Unlike the semi-classical and flavor source
terms, there is no clear physical explanation for its origin. The VIA
source crucially depends on thermal corrections, in particular on the
thermal width, making it distinct from the other sources. Flavor
dynamics is essential as well, as for example in a set-up with a left-
and right-handed fermion, the left-handed source term is minus the
right-handed source. Note that the use of the term `flavor' is
somewhat cavalier in this context, and used to distinguish different
chiralities. We nevertheless stick to this terminology to highlight
the similarities between a bosonic and fermionic source term; in both
cases the source hinges on a misalignment between the 
propagating state and the state that interacts with the plasma. 
The VIA source already arises at leading
order in the derivative expansion, which treats the bubble wall
background as evolving slowly compared to all other relevant
time scales. This explains why it generates larger asymmetries
compared to other theoretical approaches \cite{Cline:2020jre,Cline:2021dkf}.

The authors of Ref.~\cite{Carena:2000id} tried to avoid the VIA
expansion, but their approach is not based on a first principle
derivation, but rather on a phenomenological method, including a
phenomenological decay width, following \cite{Huet:1995sh,
  Riotto:1998zb}.  In Ref.~\cite{Cline:2020jre} it was noted that for
degenerate masses the source term diverges in the zero width
limit. However, the approximations made to derive the VIA source
already break down before this limit is reached \cite{Postma:2021zux}.
Ref.~\cite{Kainulainen:2021oqs} highlighted technical issues with the
VIA source, such as omitted pole contributions and renormalization
issues. While these indeed should be properly addressed, they do not
directly invalidate the derivation, as the renormalization issue does
not arise in a bosonic system, and the extra pole contributions \emph{can} be
included. More fundamentally compromising for the VIA approach,
  Ref.~\cite{Kainulainen:2021oqs} also showed that VIA-like sources
  vanish at all orders in the derivative expansion if the interactions
  with the thermal bath are mediated by QCD. This result can be
  understood because the VIA source arises from the axial current equation,
  which vanishes in a vector-like theory. Indeed, all equations can be
  written in terms of the phase space densities of Dirac fermions, and
  thus no VIA source that is equal but opposite in sign for the left- and right-handed
  fermions is generated.  In this paper we allow
  for chiral interactions with the bath which can yield a non-zero
  axial current, in line with the implicit  assumption
  usually made in the VIA literature.

Although the VIA-method already appeared in \cite{Huet:1994jb,Huet:1995mm,Huet:1995sh}, the source term
was first derived in the closed-time-path formalism in
\cite{Riotto:1995hh, Riotto:1998zb}, and subsequently generalized to
include a CP-conserving relaxation rate in \cite{Lee:2004we}. Both the
CP-violating and -conserving source terms are derived from the
Schwinger-Dyson equations by expanding the self-energies to second
order in the VIA expansion. This results in a self-energy diagram that
is non-local \cite{Kainulainen:2021oqs} and not
one-particle-irreducible (1PI) 
\cite{Postma:2021zux}.
This approach makes it hard to 
  count  diagrams correctly, which is an obstacle to resumming all
  self-energy insertions.
 The source term is derived from the collision term in
the transport equations, which is evaluated with a second order
self-energy and zeroth order Green's function.  In
\cite{Postma:2021zux} it was demonstrated that the same source term
can be obtained from the mass commutator in the transport equations, if
the self-energy is evaluated at first order -- this is a 1PI diagram
-- and the Green's function is computed to first order in VIA as
well. This equivalence strongly suggest the technical issues noted
  in \cite{Kainulainen:2021oqs} do not arise from the non-local nature
  of the self-energy diagrams appearing in the standard derivation.

In this paper, we will follow the 
approach of Ref.~\cite{Postma:2021zux}  and only include 1PI, equilibrium,
self-energy corrections, which is equivalent to treating the
space-time dependent part of the mass matrix as a perturbation. The
Green's functions at any given order in VIA then follow from the
Schwinger-Dyson equations (or in Wigner space, from the Kadanoff-Baym
equations).  The Green's functions can be determined perturbatively,
but remarkably, the full set of constraint equations can be solved
exactly as well. The VIA source term is then derived both
perturbatively as well as fully resummed over all mass insertions. We
find that the source term actually gets two contributions, from
the mass-commutator and from the collision term in the transport
equations. 
The cancellation happens order by
order in VIA, and also for the fully resummed case.  There is thus no
VIA source term at leading order in the derivative expansion.
 This
 cancellation was missed before as the original non-1PI approach
 obscures this \cite{Riotto:1995hh, Riotto:1998zb,Lee:2004we}, and
 \cite{Postma:2021zux} only considered the mass-commutator but not the
 collision term.

When something cancels exactly, there usually is a deeper
reason. Could we actually have foreseen our results?  We think we
could have.  In VIA the Green's functions are derived at leading order
in the derivative expansion, and are thus indistinguishable from the
Green's functions in a constant background. The source term is also
derived at leading order in the derivative expansion.\footnote{To be
  precise, at leading order in the expansion of the diamond operator
  in the Kadanoff-Baym equations.  In the standard VIA approach, in
  which this source does not cancel, the integrand is then
  subsequently expanded in something similar to a derivative expansion
  to extract the dominant CP-violating part -- but this is not
  equivalent to the diamond expansion.} In the limit of a constant
background the source should vanish, as equilibrium is restored.  But
this is the same limit as taking the zeroth order in the derivative
expansion, and one should thus have expected a zero source.

This paper is organized as follows. Section~\ref{sec:formalism} introduces the Kadanoff-Baym equations and the Green's functions that describe the out-of-equilibrium plasma. In Sections~\ref{s:bosons} and \ref{s:fermions} we solve the Green's functions with and without VIA expansion for bosons and fermions respectively. We also derive the corresponding source term and demonstrate that they are zero. We conclude in Section~\ref{s:conclusion}.


\section{CTP formalism and Green's functions}\label{sec:formalism}

The closed-time path (CTP), or in-in formalism, provides a
first-principle description in terms of Green's functions\footnote{We
  will use the terms `Green's function' and `propagator'
  interchangeably.} of a system out-of-equilibrium.  The
Schwinger-Dyson equations can be split into hermitian and anti-hermitian
parts, usually referred to as the constraint and kinetic equations.
The constraint equations provide the spectral information of the
system, whereas the kinetic equation determines the dynamical
evolution.  In this section we list the results and relations relevant
for the discussion of the source terms for electroweak
baryogenesis. This section also serves to set the notation. More details and
derivations can be found in e.g. \cite{Riotto:1998zb,
  Prokopec:2003pj,Cirigliano:2011di,Cirigliano:2009yt}. We use the
same conventions as \cite{Prokopec:2003pj}.

The bosonic and fermionic CTP Green's functions are, respectively,
\be
i\Delta (u,v)
 = \langle \Omega| T_{\cal C} \, \left[ \phi(u) 
 \phi^\dagger (v) \right]|\Omega\rangle, \qquad
  iS_{\alpha\beta} (u,v)
 = \langle \Omega| T_{\cal C} \, \left[ \psi_\alpha(u) \bar
 \psi_\beta (v) \right] |\Omega\rangle,
\ee
with $T_{\cal C}$ denoting time-ordering along the contour ${\cal C}$
which runs from initial time $t_0 = -\infty$ to time $t$ and back. The
contour can thus be split into a forward- and backward-going branch, and fields are
labeled by $a=\pm$, depending on the branch they lie on.  We define
$G^{ab} =\{\Delta^{ab},S^{ab}\}$ for bosons and fermions
respectively, and introduce the notation
$G^t = G^{++}, \, G^{\bar t} =G^{--},\, G^> = G^{-+}, \, G^< =
G^{+-}$.

The retarded and advanced
propagators are
\begin{align}
  G^r &\equiv G^t - G^< = G^> - G^{\bar t} ,&G^a &\equiv G^t - G^>  =  G^< - G^{\bar t},
\label{Gdefs1}
\end{align}
from which it also follows that $G^t +G^{\bar t} = G^>+ G^<$. Throughout we will label $G^\lambda$ with $\lambda =
>,\, <$ for the so-called Wightman functions, and $G^\alpha$ with $\alpha =
r,\, a$ for the retarded and advanced propagators. 
The hermiticity properties of the Wightman functions are
$(i\Delta^\lambda(u,v))^\dagger = i\Delta^\lambda(v,u)$ and
$(\gamma^0 iS^\lambda(u,v))^\dagger = \gamma^0 iS^\lambda(v,u)$ for
bosons and fermions respectively.
The retarded/advanced propagators can be split into hermitian and
anti-hermitian parts via
\be
G^{r,a} = G^h \mp i G^{\cal A} \;\;\; {\rm with} \;\;\;
G^h \equiv \frac 12 (G^r+G^a) 
\;\; \; \& \;\; \; G^{\cal A} \equiv \frac {1}{2i} (G^a-G^r),
\label{Gdefs2}
\ee
with the upper (lower) sign for the retarded (advanced) propagator in the first expression.
In thermal equilibrium, the Wightman functions can
be written in terms of the spectral function\footnote{The fermionic Wightman functions can be projected onto a
  helicity basis, and \cref{Grho} then applies to the projected
  Wightman functions, see \cref{WightmanP1,WightmanP2} for the exact
  expressions for fermions. }
\be
G^\lambda = \rho g^\lambda, \quad {\rm with}\quad
g^> = (1+ s n_s) \;\;\; \& \;\;\; g^< = s n_s \quad  {\rm and}
\quad \rho \equiv (G^r-G^a), 
\label{Grho}
\ee
with $\rho$ the spectral function. Here $s=1$ for bosons and $s=-1$ for fermions, and
$n_s = (\e^{(E-\mu)/T}-s)^{-1}$ is the Bose-Einstein or Fermi-Dirac
distribution for bosons and fermions, respectively. 

The temperature-dependent self-energies arise from loop corrections in
the plasma background. To emphasize the similarity between the bosonic
and fermionic system, we will denote the self-energies for both by
$\Pi^{ab}$, rather than using the more conventional 
$\Sigma^{ab}$ notation for the fermionic self-energy.  When approximated by
their thermal equilibrium expressions, the self-energies $ \Pi^{ab}$
satisfy the same relations as the Green functions $G^{ab}$ listed
above. In particular, the advanced/retarded self-energy can be split
into real and imaginary parts $ \Pi^{r,a} = \Pi^h \mp i \Pi^{\cal A}$,
with $\Pi^h$ containing the thermal mass corrections and
$ \Pi^{\cal A}$ the thermal widths.


\subsection{Kadanoff-Baym equations}

The position-space Schwinger-Dyson equations can be written in the
form \cite{Riotto:1998zb,Prokopec:2003pj}:
\begin{align}
	K(u)G^{ab}(u,v) &= a  \delta_{ab} \delta^4(u-v) + \sum_c c
                          \int d^4 w \, \Pi^{ac}(u,w) G^{cb}(w,v),
                          \label{eq_SD}
\end{align}
with $a,b,c= \pm$.  For bosons $K(u) = -(\partial^2_u + M_0(u)^2)$ is
the Klein-Gordon operator with $M_0^2$ the tree-level mass matrix, and
for fermions $K(u)= (i\slashed{\partial}_u- M_0(u))$ is the Dirac
operator with $ M_0(u)$ the tree-level Dirac mass.

For thick bubble walls, one can perform a gradient expansion in the
slowly varying bubble background. It is then useful to define the
relative and center-of-mass coordinates $ r = u-v$ and
$x= \frac12(u+v)$, and to make a Wigner transform, which is defined as the Fourier
transform with respect to the relative coordinate:
\be
G(k,x) = \int \dd^4 (u-v) \, \e^{ik.(u-v)} G(u,v) =\int \dd^4 r \,
\e^{ik.r} G(x+\frac12 r,x- \frac12 r).
\ee
The gradient expansion in the slowly varying background is then
equivalent to an expansion in the diamond
operator, which is defined as
\be
\diamond \big(A(k,x) B(k,x) \big)= \frac12 \big(\partial_x A(k,x) \cdot\partial_k B(k,x) -\partial_k A(k,x) \cdot \partial_x B(k,x)\big).
\ee

Wigner transforming the hermitian and anti-hermitian part of the Schwinger-Dyson
equation (\ref{eq_SD}) gives the constraint and kinetic equations, known as the
Kadanoff-Baym (KB) equations. For bosons this gives
\begin{align}
 ( k^2 -\frac14 \partial_x^2)G^{ab} &= a  \delta_{ab}  \otimes 1+
                                      \frac12\e^{-i\diamond}\Big(  \{M_0^2,
                                      G^{ab}\}
                                      +\sum_c c (\Pi^{ac}G^{cb}+G^{ac}\Pi^{cb})\Big),
                                      \nn \\
  2 ik\cdot \partial_xG^{ab} &= 
                                     \e^{-i\diamond}\Big(  [M_0^2,
                                      G^{ab}]
                                +\sum_c c (\Pi^{ac}G^{cb}- G^{ac}\Pi^{cb})\Big).
 \label{KBab}
 \end{align}              
Note that $G^{ab}, \Pi^{ab}, M_0^2$ and $1$ are $n\times n$-matrices
in flavor space with $n$ complex bosonic degrees of freedom. For fermions we get
\begin{align}
 \frac12\{\slashed{k},G^{ab}\} &= a  \delta_{ab}  \otimes 1+
                                      \frac12\e^{-i\diamond}\Big(
                                 \{ M_0,
                                      G^{ab}\}
                                      +\sum_c c (\Pi^{ac}G^{cb}+G^{ac}\Pi^{cb})\Big),
                                      \nn \\
\frac{i}2\{\slashed{\partial}_x,G^{ab}\} &= 
                                     \e^{-i\diamond}\Big(   [ M_0,
                                      G^{ab}]
                                +\sum_c c (\Pi^{ac}G^{cb}- G^{ac}\Pi^{cb})\Big).
 \label{KBabF}
 \end{align}              
For a single Dirac fermion $G^{ab}, \Pi^{ab} , \delta M$ and $1$ are
matrices in Dirac space; projecting onto left and right chiralities,
this can be viewed as $2\times 2$ in `flavor' space.

\subsection{Retarded/Advanced propagator and thermal corrections}

Introducing the matrices
\be
\tilde G =\(  \begin{array}{cc} G^{t} &- G^{<} \\
         G^{>} & -G^{\bar t} \end{array} \), \qquad
\tilde \Pi =\(  \begin{array}{cc} \Pi^{t} & - \Pi^{<} \\
                         \Pi^{>} & - \Pi^{\bar t} \end{array} \),
 \label{Gtilde}
\ee
the Schwinger-Dyson \cref{eq_SD} takes the simple form \cite{Riotto:1998zb}
\begin{align}
	K(u)\tilde G(u,v) &= \delta^4(u-v) + 
                          \int d^4 w \, \tilde \Pi(u,w) \tilde G(w,v).
\end{align}
For a single flavor system in thermal equilibrium the self-energies
and Wightman functions satisfy the KMS relations
\be
(1+s n_s) G^<=  s n_s G^> ,\quad (1+s n_s)  \Pi^<=  s n_s \Pi^> .
\label{KMS}
\ee
We note that in the presence of flavor mixing the equilibrium
conditions are no longer of the simple KMS form \cref{KMS}. We will
return to this point in \cref{s:constraint_B}.

We can diagonalize the propagator and self-energy matrices by an
orthogonal matrix $V$ (with $V V^{-1} =1$)
\be
 V \tilde G
V^{-1}  =\( \begin{array}{cc} G^{r} & 0 
     \\
              0 & G^{a} \end{array} \), \quad
 V \tilde \Pi
V^{-1}  =\( \begin{array}{cc} \Pi^{r} & 0 
     \\
 0 & \Pi^{a} \end{array} \), \quad{\rm with} \;\;
V =\(  \begin{array}{cc}(1+sn_s)& -s n_s \\
1& -1 \end{array} \) .
\label{V}
\ee                   
The diagonal entries $G^{r,a}$ and $\Pi^{r,a}$ are the
retarded/advanced propagators and self-energies respectively.
Eqs.~(\ref{Gtilde}) and (\ref{V}) contain the same information as the relation between
the Green's functions in \cref{Gdefs1,Grho}.  Using this rotation, the
Schwinger-Dyson equations, and thus also the Wigner space KB
equations, can be diagonalized.

\section{Bosons}
\label{s:bosons}

In this section we will derive the VIA source for a set-up with two
flavors and an off-diagonal spacetime-dependent mass.  We assume that
the interactions with the bath are flavor-diagonal and are different
for the two flavors.
The results can
be generalized to other bosonic systems as well. After discussing the
set-up and the thermal corrections, we will first solve the constraint
equation and derive the VIA source perturbatively, expanding in the
number of mass insertions. However, the constraint equations can also
be solved exactly, without such an expansion, and we also present the corresponding
resummed results.

\subsection{Lagrangian}

For definiteness, consider a toy model with two bosonic flavors,
labeled by $L,R$, with a CP-violating interaction with the Higgs field. In terms of the flavor doublet $\phi=1/\sqrt2( \phi_L \; \phi_R)^T$, the
quadratic Lagrangian in the bubble wall background is
\be
\L^{(2)} = (\partial_\mu \phi)^\dagger (\partial^\mu \phi)
-\phi^\dagger M_0^2 \phi, \qquad
M_0^2(v) =
\(
\begin{array}{cc}
  m_{LL,0}^2 & m_{LR}^2(v) \\
 m_{RL}^2(v)  & m_{RR,0}^2
\end{array}
\).
\label{M_boson}
\ee
The subscript $0$ indicates that this is the zero temperature mass
matrix. The field-independent diagonal mass terms will receive finite
temperature corrections, which are diagonal in flavor space for a
flavor-diagonal coupling to the heat bath. The off-diagonal masses
depend on the background Higgs vev $v$, which is spacetime dependent, and violates CP in the bubble wall
background.
The vev insertion approximation consists of treating
$m^2_{LR} = (m^2_{RL})^*$ as a perturbation, and expanding the
KB-equations in this small quantity.

\subsection{KB equations}
\label{s:KB_B}

The constraint equations for the Wightman functions $G^\lambda$ with
$\lambda = >,\,<$ and time-ordered propagator $G^t$ follow from  \cref{KBab} 
\begin{align}
 \(k^2-\frac14 \partial_x^2 \) G^\lambda&= \frac12\e^{-i\diamond}\(
\{M^2 , G^\lambda\} +\{\Pi^\lambda , G^h\} + \frac12([\Pi^> ,
                                          G^<] -[\Pi^< , G^>] ) \),
    \nn
  \\
 \(k^2-\frac14 \partial_x^2 \) G^t&= 1+\frac12\e^{-i\diamond}\( 
                                     \{M^2+ \Pi^t-\Pi^h , G^t\} -\Pi^<
                                    G^> -G^< \Pi^>\).
    \label{KB_boson}
\end{align}
 Here we rewrote the right-hand side of the first equation using the
following relations (and the equivalent ones with $\Pi$ and $G$ flipped)
\begin{align}
\Pi^> G^t
-\Pi^{\bar t} G^> &= \Pi^h G^> + \Pi^> G^h +\frac12(\Pi^>G^< -\Pi^<
G^>), \nn \\
\Pi^t G^<
-\Pi^< G^{\bar t}  &= \Pi^h G^< + \Pi^< G^h +\frac12(\Pi^>G^< -\Pi^<
G^>).
\end{align}
We have further defined $M^2 = M_0^2 + \Pi^h$, where $M_0^2$ is the
tree-level mass matrix, and $\Pi^h$ the thermal mass, as shown in the
next subsection.  Note that since the system of equations only depends
on $M^2$, it is fully equivalent to include the space-time dependent
mass in $M_0^2$ and expand the mass matrix, or to include it in the
interaction Lagrangian and add the 1PI diagram with one mass insertion
to $\Pi^h$ \cite{Postma:2021zux}. Here we follow the former
approach.

The kinetic equation for the Wightman functions is
\begin{align}
 2 ik\cdot \partial_x G^\lambda&  = \e^{-i\diamond}\Big(
[M^2, G^\lambda]+ [\Pi^\lambda,
                                           G^h] +
                                          \frac12\(\{\Pi^> , G^<\}-\{\Pi^< ,
G^>\}\) \Big) .
    \label{KB_boson}
\end{align}
Integrating the left-hand side over four-momenta and adding up the two
different Wightman functions gives \cite{Riotto:1998zb} 
%
\be
 \frac 1 2\partial_\mu \int \frac{\dd^4 k }{(2\pi)^4} ik^\mu \(G^>(k,x)+G^<(k,x) \)
 =- i\langle\phi^\dagger(x)\overset{\leftrightarrow}{\partial^\mu} \phi(x) \rangle =
- \partial_\mu \langle J^\mu(x)\rangle.
 \ee
We can then identify the source term for the left- and right-handed
flavors from the right-hand side of the kinetic equation. The
integration over 4-momenta only picks out the leading order term in the
diamond expansion, which amounts to setting $\e^{-i \diamond}
=1$. Explicitly, the left-handed source is
\begin{align}
S_{LL} &= -\int\frac{\dd^4 k }{(2\pi)^4} \(  [M^2, G^>+G^<]+ [\Pi^>+\Pi^<,
                                           G^h] +
                                       \{\Pi^> , G^<\}-\{\Pi^< ,
                                       G^>\} \)_{LL} \nn \\
    &\equiv -\int\frac{\dd^4 k }{(2\pi)^4} \bar S_{LL},
      \label{Sbar_B}
\end{align}                                        
where the label $LL$ refers to the $(1,1)$-component of the resulting source matrix.
We introduced the notation $\bar S_{LL}$ for the integrand for future
convenience.

\subsection{Thermal corrections}
\label{s:thermal_B}

Let's first consider the solutions of the constraint equation for a
flavor-diagonal mass matrix and self-energy, and see how the thermal
self-energies affect the solution. This would be the starting point in
the vev-insertion approximation, where at leading order the
off-diagonal mass terms are set to zero. The constraint equations for
different flavors decouple, and we can focus on a single flavor. We
work at leading order in the derivative expansion and set
$\e^{-i \diamond}=1$.

If the thermal plasma is close to equilibrium,  to first
approximation the thermal self-energies and propagators satisfy the KMS
relations \cref{KMS}. We can then rotate the KB equations to the basis
of retarded and advanced propagators using \cref{V}:
\be
k^2 G^\alpha - \frac12 \{M_0^2+\Pi^\alpha,G^\alpha\} = (k^2 - M_0^2 -
\Pi^\alpha) G^\alpha = 1,
\label{Gr_result}
\ee
with $\alpha =r,a$.
The self-energies for each flavor can be rewritten in terms of the hermitian and
anti-hermitian parts \cref{Gdefs2}, which are related
to the thermal mass $m_T$ and width $\Gamma_T$ respectively,  via \cite{Greiner:1998vd, Riotto:1998zb,Henning:1996dv}
\begin{align}
 \Pi^\alpha = \Pi^h \mp i \Pi^{\cal A}, \quad {\rm with} \quad \Pi^h =
  m^2_T \;\;\; \&\;\;\;\Pi^{\cal A} = 2k_0 \Gamma, 
\label{Pi_boson_T}                                                      
\end{align}
with the upper (lower) sign for the retarded (advanced) propagator. The sign (and factor 2) in the bosonic case are taken to obtain the same result
as in \cite{Riotto:1995hh,Lee:2004we} for the spectral
function. 
The constraint equation for a single flavor is solved by
\begin{align}
(G^\alpha)^{-1} &= k^2 -m_0^2 -  \Pi^h_T \pm i \Pi_T^{\cal A}.
 \label{Gra_single}              
\end{align}
The
Wightman functions are given by \cref{Grho} in terms of the spectral
function 
\be
\rho =G^r -G^a= \frac{-2 i \Pi^{\cal A}}{ (k^2 - m_0^2 - \Pi^h)^2 + (\Pi^{\cal A})^2 }
= \frac{\gamma}{ (k^2 - m^2)^2 -\gamma^2/4 },
\label{rho1}
\ee
with $m^2 =m_0^2+ \Pi^h$ the tree-level plus thermal mass, and
$ \gamma = -4 i k_0 \Gamma $.

\subsection{Vev-insertion approximation}
\label{s:VIA_B}

Just as in the standard VIA approach, we work at leading order in the
derivative expansion and set $\e^{-i\diamond} =1$ in the KB
equations. In VIA the off-diagonal mass term is treated as a
perturbation, and one can solve the KB equations perturbatively as a
series in the number of `mass insertions', i.e. vev insertions
\cite{Riotto:1995hh,Riotto:1997vy,Riotto:1998zb,Lee:2004we}.  The
solution of the constraint equations can be substituted into the kinetic
equation to get the dynamics of the system and derive the source term
\cref{Sbar_B}. The source term is calculated at 2nd order in VIA, which
requires solving the constraint equation to 2nd order as well.

\subsubsection{Constraint equation}
\label{s:constraint_B}

The mass matrix can be expanded as $M^2 =M_d^2 + \delta M^2$, with
$M_d$ containing the diagonal tree-level plus thermal masses, and $\delta M^2$
the off-diagonal and field-dependent masses. The constraint equations
at   leading order in VIA -- i.e. setting $\delta M^2=0$ -- are
\begin{align}
 k^2G_{(0)}^\lambda&= \frac12\Big(
\{M_d^2 , G_{(0)}^\lambda\} +\{\Pi^\lambda , G_{(0)}^h\} + \frac12([\Pi^> ,
                                          G_{(0)}^<] -[\Pi^< , G_{(0)}^>] ) \Big),
    \nn
  \\
k^2G_{(0)}^t&= 1+ \frac12\Big(
                                     \{M_d^2+ \Pi^t-\Pi^h , G_{(0)}^t\} -\Pi^<
              G_{(0)}^> -G_{(0)}^< \Pi^>\Big),
\label{constr_B_0}
\end{align}
where we have set $\e^{-i\diamond} =1$. We have also dropped the
$\frac14 \partial_x^2 G_{(0)}^\lambda$ term, as this term likewise
leads to derivatives of the background dependent mass, which is
assumed to be sub-leading in the derivative expansion.  This is not different
from the standard assumption in the VIA computation.
The thermal corrections arise from (approximately) thermal equilibrium
physics, and we assume that the self-energies satisfy KMS relations,
and are flavor diagonal:
\be \Pi_{IJ}^\lambda =g^\lambda_{II} \gamma_I \delta_{IJ},
\qquad \gamma_I = -4ik_0 \Gamma_I.
\label{gamma}
\ee
with $I,J =L,R$ and  $g^\lambda$ given in \cref{Grho}.
 The off-diagonal
self-energies vanish. 

The thermal self-energies are given in \cref{gamma}.  The 0th order KB
equations are solved by Wightman functions that are flavor diagonal
and that satisfy the KMS relations
\be
G_{(0),IJ}^\lambda = g^\lambda_{II}\rho_{(0),I} \delta_{IJ}.
\label{Wightman0}
\ee
Substituting this and \cref{gamma} into the leading order constraint
equations (\ref{constr_B_0}), one can solve for the spectral
function\footnote{At the pole $k_0 \approx \omega_I=\sqrt{|\vec k|^2+m_I^2}$ and we can write
  the spectral function in the form
\begin{align}
  \rho_{(0),I} 
 &  = \frac{1}{2 \omega_I} \[ \(\frac{1}{k^0 - \omega_I + i  \Gamma_I}  - 
  \frac{1}{k^0 + \omega_I + i  \Gamma_I}\) - \(\frac{1}{k^0 - \omega_I - i  \Gamma_I}  - 
   \frac{1}{k^0 + \omega_I- i  \Gamma_I}\) \]. \nn 
\end{align}
 This is the standard
form of the zeroth order propagator used in the VIA approach
\cite{Lee:2004we}.}
\begin{align}
  \rho_{(0),I} &= \frac{\gamma_I}{(k^2-m_I^2)^2 - \gamma_I^2/4} =
                 \frac{\gamma_I}{D_{I} D^*_{I}},
             \label{rho} 
\end{align}
with $D_{I} = k^2-m_I^2- \gamma_I/2$ and $D^*_{I} = k^2-m_I^2+ \gamma_I/2$.
The spectral function
agrees with the result derived in the basis of retarded and advanced
propagators in thermal equilibrium \cref{rho1}. 
The solution for the time-ordered Green's function is
\be
G_{(0),IJ}^t =\[ n_I (G^r - G^a) + G^r \] \delta_{IJ}=
\rho_{(0),I} \[ n_I 
  +\frac{D^*_I}{\gamma_I}\]\delta_{IJ},
\ee
where the first expression can also be derived from \cref{V}, and we
used that $G^r_I =D_{I}^{-1}$ and $G_I^a =(D_{I}^*)^{-1}$ in the
right-most expression.  The anti-time-ordered Green's function is
\be
G_{(0),IJ}^{\bar t} =\[ n_I (G^r - G^a) - G^a \] \delta_{IJ}=
\rho_{(0),I} \[ n_I 
  -\frac{D_I}{\gamma_I}\]\delta_{IJ}.
\ee

The propagators at 1st and 2nd order in VIA can be expressed in terms of the
zeroth order propagators by means of the Schwinger-Dyson equation
\begin{align}
G^{ab}_{(1),IJ} &=  \sum_c c\, G^{ac}_{(0),II} (\delta M^2)_{IJ}
                  G^{cb}_{(0),JJ} , \nn \\
  G^{ab}_{(2),II} &=  \sum_{cd} cd\, G^{ac}_{(0),II} (\delta M^2)_{IJ}
                    G^{cd}_{(0),JJ} (\delta M^2)_{JI}G^{db}_{(0),II} 
                    \label{G12},
\end{align}
with $J \neq I$.
The first and second order propagators are flavor-off-diagonal and
-diagonal respectively. 

\subsubsection{Kinetic equation}
\label{s:kinetic_B}

For notational convenience we will drop the (0) subscript on the zeroth
order Green's functions and spectral function in this subsection.  The
higher order Green's functions are labeled by a subscript, and there
should thus be no source of confusion.

At zeroth order the source term vanishes as the zeroth order propagators
are diagonal and therefore the commutator with the mass vanishes. As the zeroth
order propagators satisfy the KMS relation the collision term also
vanishes.  At first order the diagonal entries still vanish by virtue
of the KMS relations for the zeroth order propagators. To show that the
off-diagonal source vanishes as well is a bit more non-trivial, but
straightforward. Here we focus on the diagonal source terms at 2nd
order, which is the usual order at which the VIA source is derived,
given by \cref{Sbar_B}
\begin{align}
\bar S^{(2)} &  = 
[\delta M^2,  (G_{(1)}^>+G_{(1)}^<)]+[M_d^2,  (G_{(2)}^>+G_{(2)}^<)]
                                            + [\Pi^>+\Pi^<,
                                           G^h_{(2)}]  \nn \\ &+
                                          \(\{\Pi^> , G_{(2)}^<\}-\{\Pi^< ,
                                                      G_{(2)}^>\}\) .
                                                      \label{S2}
\end{align}
Only the first and last terms on the right-hand-side contribute to the
diagonal source term $(\bar S^{(2)})_{II}$. The higher order Green's
functions can be expressed in terms of the zeroth order Green's functions using
\cref{G12}.

The contribution of the first term in \cref{S2} to the
$LL$-component of the source is
\begin{align}
 \bar S^{(2)}_{M,LL} & \equiv [\delta M^2,  (G_{(1)}^>+G_{(1)}^<)]
=m^2_{LR} (G^>_{(1),RL}+G^<_{(1),RL})-
                  (G^>_{(1),LR}+G^<_{(1),LR}) m^2_{RL} \nn \\
  &= | m_{LR}|^4 \( (G^>-G^<) _{RR} (G^t+G^{\bar
               t})_{LL}-(G^>-G^<) _{LL} (G^t+G^{\bar
               t})_{RR} \)\nn \\
  &=
   |m_{LR}|^4 \rho_L\rho_R \[  (2n_L-1) -  (2n_R-1) \]
   =
    2 |m_{LR}|^4 \rho_L \rho_R (n_L-n_R),
    \label{S2M_B}
\end{align}  
where we used that $(G^t +G^{\bar
  t})_{II}= (2n_I-1) \rho_I$ and $(G^> -G^<)_{II} =
\rho_I$. 
The collision term contributes
\begin{align}\label{S2M_C}
\bar S^{(2)}_{C,LL} &\equiv
          \left(  \{\Pi^> , G_{(2)}^<\}-\{\Pi^< ,
                                                      G_{(2)}^>\} \right)_{LL}\
             \\
 &=2 | m_{LR}|^4\, \Big[ (\Pi^>_{LL} G^<_{LL} -G^>_{LL}\Pi^<_{LL}) ( G^{\bar t}_{LL}
                G^{\bar t}_{RR} +G^t_{LL}G^t_{RR}) 
                 \nn \\ 
                     &\hspace{1.5cm}
                       +(\Pi^<_{LL} G^>_{LL} G^<_{RR} G^>_{LL} -\Pi^>_{LL}
    G^<_{LL} G^>_{RR} G^<_{LL})
                +
  (\Pi^<_{LL}G^>_{RR} -\Pi^>_{LL}G^<_{RR}  )
          G^{\bar t}_{LL}   G^t_{LL}\Big].\nn
\end{align}
The term on the 2nd line vanishes by virtue of the KMS relation for the zeroth
order propagator and self-energy. The first term on the last line becomes
becomes 
\begin{align}
  {\rm term}_1& \equiv
(\Pi^<_{LL} G^>_{LL} G^<_{RR} G^>_{LL} -\Pi^>_{LL}
    G^<_{LL} G^>_{RR} G^<_{LL})
                \nn \\
&= \gamma_L \rho_L^2 \rho_R  n_L (1+n_L)  \( n_R (1+n_L)
         - n_L (1+n_R)  \) 
         \nn \\ &= -\gamma_L  \rho_L^2 \rho_R n_L (1+n_L)(n_L-n_R).
\end{align}
The second term on the last line equals
\begin{align}
{\rm term}_2 &\equiv (\Pi^<_{LL}G^>_{RR} -\Pi^>_{LL}G^<_{RR}  )
          G^{\bar t}_{LL}   G^t_{LL}\nn \\
    & = \gamma_L  \rho_L^2 \rho_R \( n_L -n_R\)
      \(n_L +  \frac{D^*_{L}}{ \gamma_L} \) \(n_L -
             \frac{D_{L}}{\gamma_L} \).
\end{align}
Adding them up we find
\begin{align}
({\rm term}_1+{\rm term}_2)& = - \bar \gamma_L  (\rho_L^+)^2 \rho_R^-(n_L-n_R)\[
                           n_L (1-n_L)+\(n_L +
             \frac{D^*_{L,+}}{\bar \gamma_L} \) \(n_L -
             \frac{D_{L,+}}{\bar \gamma_L} \)  \]
\nn \\ & = -\gamma_L  \rho_L^2 \rho_R(n_L-n_R)
         \frac{D^*_{L}D_{L}}{ \gamma_L^2}
        = -\rho_L \rho_R(n_L-n_R),\label{SLL_C}
\end{align}
where we used that $D_{L}-D^*_{L}+\gamma_L=0$.
The final result is then
\begin{align}
 \bar  S_{LL}^{(2)} &=\bar S^{(2)}_{M,LL} +\bar S^{(2)}_{C,LL} = 
                         2 |m_{LR}|^4 \rho_L\rho_R (n_L-n_R) \nn
                     -  2 | m_{LR}|^2  \rho_L\rho_R  (n_L-n_R) 
                      =0.
\end{align}                 
The VIA source thus cancels to 2nd order in the VIA expansion.  Likewise,
the right-handed source term and all off-diagonal contributions to the
source term vanish.

\subsection{Comparison with usual derivation of the VIA source}

Let's compare our results with the standard derivation of the VIA
source found in the literature
\cite{Riotto:1995hh,Lee:2004we}. In this approach the self-energy is
expanded to 2nd order
\be \Pi^\lambda_{II,(2)} = -(\delta M^2)_{IJ} G^\lambda_{JJ}(\delta
M^2)_{JI},
\ee
and Wightman functions are taken at leading order. 
The diagonal source
originates from the collision term as all other terms in the kinetic
equation vanish
\begin{align}
\bar S_{LL}^{(2)}\big|_{\rm usual} &  = 
\{\Pi_{LL,(2)}^> , G_{LL}^<\}-\{\Pi_{(LL),(2)}^< ,
                                                      G_{LL}^>\} \nn \\ &=-|m_{LR}|^4
                                     \(\{ G_{RR}^>,G_{LL}^<\}  -\{
                                     G_{RR}^<,G_{LL}^>\}\)
   =-|m_{LR}|^4 \rho_L \rho_R (n_L-n_R). \label{eqn:colsource}
\end{align}
This is the same expression for the collision term as in our approach
derived above \cref{SLL_C}.  To bring this into more standard VIA notation
we perform an inverse Wigner transform on the first line in
\cref{eqn:colsource} above to evaluate the source \cref{Sbar_B} in position space:
\begin{eqnarray}
  S_{LL}^{(2)}(x) 
  &=& -2 \int d^4 y {\rm Re} \left[ m_{LR}^2 (x) G_{RR}^<(x,y)
      m_{RL}^2 (y) G_{LL}^> (y,x) \right. \\ && \left. \hspace{1.95cm} -
 m_{RL}^2(x)G_{RR}^>x,y)m_{LR}^2 (y) G_{LL}^<(y,x)\right]
                                                          \nn \\
  &=& - 2 \int d^4 y \, \Big( {\rm Re}\left[
                        g(x,y)+g(y,x) \right]  {\rm Re}\left[ G_{RR}^<(x,y) G_{LL}^> (y,x) - G_{RR}^>(x,y)G^< _{LL}(y,x)  \right] \nonumber \\ 
&& \hspace{1.6cm}+    {\rm Im}\left[
                        g(x,y)+g(y,x) \right] 
{\rm Im} \left[ G_{RR}^<(x,y) G_{LL}^> (y,x) - G_{RR}^>(x,y)G^<
   _{LL}(y,x)  \right] \Big) ,  \nn
\end{eqnarray}
with $g(x,y) = m_{LR}^2 (x) m_{RL}^2(y)$, and where for notational
convenience we have dropped the `usual' subscript.
Now expand the masses in the integrand
\begin{equation}
m_{IJ}^2(y) = m_{IJ}^2(x) + (x-y)^\mu \partial _\mu m^2_{IJ}(x)
+{\cal O}(\partial^2).
\end{equation}
Although this expansion also assumes a slowly varying bubble wall
background, it is different from the derivative expansion, i.e. the
expansion in terms of the diamond operator, of the KB equations. The
leading order term of this mass expansion gives a CP-conserving source
term, whereas from the next-to-leading term we recover the usual CP-violating (CPV) source term found in the literature:
\begin{align}
    S^{(2)}_{\rm CPV}=2 {\rm Im}\left[ m_{LR}^2 \partial _\mu m^2 _{RL} \right]\int d^4 y\, (y-x)^\mu
    & \Big( G_{RR}^<(x,y) G_{LL}^> (y,x) \nn \\ & \qquad - G^>_{RR}(x,y) G_{LL}^<(y,x)  \Big) .
\end{align}
Hence, it was concluded that the VIA source term is non-zero. 
 We believe that this erroneous result is a consequence of the non-1PI
 approach, which is not a systematic perturbative expansion, and
 obscures the correct counting of diagrams.
Indeed, our more systematic approach of the previous subsections shows that if all contributions
 are carefully taken into account the source term cancels.

In \cite{Postma:2019scv} the self-energy was calculated to
next-to-leading order, that is, to 4th order in the VIA
expansion.
 This calculation thus likewise works with non 1PI-diagrams,
 and thus suffers from the same issues as the 2nd order calculation.
As
we will see in the next subsection, the VIA source cancels to all
orders in the vev-insertion expansion.

\subsection{Resummed Green's functions and source term}
\label{s:resummed_B}

With the thermal equilibrium approximation for the thermal energies in
\cref{gamma}, the KB equations at leading order in the derivative
expansion can be solved exactly for the off-diagonal mass matrix, and
there is no need to actually perform a vev-insertion expansion. The
equations to solve are
\begin{align}
 k^2G^\lambda&= \frac12\(
\{M^2 , G^\lambda\} +\{\Pi^\lambda , G^h\} + \frac12([\Pi^> ,
                                          G^<] -[\Pi^< , G^>] ) \),
    \nn
  \\
k^2G^t&= 1+ \frac12\(
                                     \{M^2+ \Pi^t-\Pi^h , G^t\} -\Pi^<
        G^> -G^< \Pi^>\),
        \label{KB_resummed_B}
\end{align}
with thermal self-energies given in \cref{gamma}.
The solutions for the Wightman functions are
\begin{align}
  G^\lambda_{LL}&= \frac{1}{\D_+ \D_-} \frac{\gamma_R \gamma_L}{\rho_R}
                  \(g^\lambda_L +g^\lambda_R \frac{\rho_R}{\gamma_L}
                  |m_{LR}|^4\),\nn \\
  G^\lambda_{LR}&=  \frac{ m_{LR}^2  }{\D_+ \D_-} \(\gamma_R g_R^\lambda
                  (k^2-m_L^2) +\gamma_L g^\lambda_L (k^2-m_R^2)
                  +\frac12 \gamma_R\gamma_L(g^\lambda_R-g^\lambda_L)\),
\label{Wresummed_B}                  
\end{align}
with $g^\lambda_I$ defined in  \cref{Grho}, $\rho_L,\rho_R$ the zeroth order spectral function \cref{rho}, and 
\be
\D_\pm = (k^2 -m_L^2 \pm \gamma_L/2) (k^2 -m_R^2 \pm \gamma_R/2) -|m_{LR}|^2.
\ee
The other flavor components can be found through $G^\lambda_{RR} = G^\lambda_{LL}|_{L
  \leftrightarrow R}$ and  $G^\lambda_{RL} = G^\lambda_{LR}|_{L
  \leftrightarrow R}$.
Note that
\be
\lim_{m^2_{LR},m^2_{RL} \to 0} \left\{ \frac{1}{\D^+ \D^-}
\frac{\gamma_R\gamma_L}{\rho_R},\,
\frac{1}{\D^+ \D^-} \frac{\gamma_R\gamma_L}{\rho_L}
\right\}=\{\rho_L,\,\rho_R\},
\ee
and we retrieve the perturbative zeroth order Green's function in this
limit.

Note that the denominators of the resummed propagators are shifted by
the off-diagonal mass squared $|m_{LR}|^2$.  Consequently, the separate
contributions from the collision term and from the mass commutator
term do not suffer from a pinch singularity, as foreseen in
\cite{Kainulainen:2021oqs}, as the divergence in the limit of equal
hermitian self-energies $\Pi^h_{LL} =\Pi^h_{RR}$ and vanishing decay
widths $\gamma_I \to 0$ is cut off.

To evaluate the off-diagonal source term one
also needs the (anti) time-ordered Green's function 
\begin{align}
G^t_{LR} &=\frac{ m_{LR}^2  }{\cal{D}_+\cal{D}_-} \Big((k^2-m_L^2) \gamma_R (n_R+\frac12)
           +(k^2-m_R^2) \gamma_L (n_L+\frac12) - |m_{LR}|^2 \nn \\
  &\hspace{2cm} + \frac14\gamma_L \gamma_R
(2n_R-2n_L+1) + (k^2-m_R^2)(k^2 -m_L^2) \Big),
\label{Gresummed_B}
\end{align}
and $G^{\bar t}_{LR} = -G^t_{LR} |_{\gamma_I \to -\gamma_I}$.

Using the explicit solutions, the source term \cref{Sbar_B}
\begin{align}
\bar S&=[M^2,  (G^>+G^<)]
                                                 + [\Pi^>+\Pi^<,
                                           G^h] +
                                          \(\{\Pi^> , G^<\}-\{\Pi^< ,
G^>\}\)  =0,
\end{align}
vanishes identically.  For the flavor-diagonal
source terms, this is due to a cancellation between the commutator with
the off-diagonal mass $\delta M^2$ and the collision term.
Explicitly, using \cref{Wresummed_B}, we find
\begin{align}
[\delta M^2,  (G^>+G^<)]_{LL} &=2 | m_{LR}|^4\frac{ \gamma_L\gamma_R (n_L-n_R)}{\D_+ \D_-} ,\nn \\
  \{\Pi^> , G^<\}_{LL}-\{\Pi^< ,
G^>\}_{LL}&=- 2 | m_{LR}|^4\frac{ \gamma_L\gamma_R (n_L-n_R)}{\D_+
            \D_-} .
            \label{B_collision}
\end{align}
And thus $\bar S_{LL}=0$ exactly, to all orders in the vev-insertion
expansion.

\subsection{Discussion: flavor dynamics}
\label{s:disc_flav}

The thermal corrections arise from model-dependent interactions with
the bath that are (approximately) described by thermal equilibrium physics. The flavor physics is only non-trivial
when the interaction and mass bases are not aligned.
 The standard VIA approach generates a
source term for the two flavors that is equal but opposite in sign,
and thus depends crucially on non-trivial flavor physics. Our choice
of the self-energy \cref{gamma} yields at leading order in the VIA
expansion a Wightman function
$G_{II}^\lambda \propto \Pi_{II}^\lambda \propto g^\lambda_{II}$ as in
\cref{Wightman0}, which is of the usual form used in the VIA
literature \cite{Riotto:1995hh, Riotto:1998zb}.  In this work it is
our aim to show that, following the assumptions made in the
literature, the VIA source cancels at leading order in the derivative
expansion. We note, though, that the cancellation of the source is
independent of the specific choice of self-energy. This can be checked
explicitly by solving the KB equations for generic (flavor-diagonal)
$\Pi^\lambda_{II}$.

At leading order in the gradient expansion the individual terms in the kinetic equation are
proportional to $(n_L -n_R)$.  Although the source term cancels at this order
in the derivative expansion, a non-zero source may still arise at
higher order. This would require non-trivial flavor dynamics,
different interactions for the flavor states, to allow for a symmetry
$n_L \neq n_R$ to build up. This is consistent with Ref. \cite{Kainulainen:2021oqs}, who did not find a VIA source at any order of the derivative expansion for fermions with vector-like bath interactions (i.e. no flavour dynamics). The fermionic case is discussed in the next section. 

Lastly, we note that the propagators at higher order in the VIA
expansion \cref{G12} and the fully resummed propagators
\cref{Wresummed_B} no longer satisfy the KMS relations as in
\cref{Wightman0} with the spectral function derived at the same order
in VIA. The reason is that at lowest order in VIA the mass and flavor
eigenstates still coincide, but at higher order this is no longer the
case. Whereas the spectral function gives information about the mass
eigenstates, the KMS relations for the self-energies are associated to
the number densities of the flavor eigenstates. This is motivated by
the rapid interactions with the bath, and to a good approximation the
bath can be regarded as a collection of particles in flavor
eigenstates.  Had $\delta M^2$ instead been flavor-diagonal, and thus
flavor and mass eigenstates aligned at all orders in VIA, then one
would have found a KMS relation \cref{Wightman0} for the Wightman
functions at all orders. The collision term in the kinetic equations
vanishes for Wightman functions and self-energies that satisfy the KMS
relations. Since this is not the case at higher order in VIA, the
collision term will contribute to the source.

\section{Fermions}\label{s:fermions}

Although the constraint and kinetic equations for bosons \cref{KBab}
and fermions \cref{KBabF} have a very similar structure, the spinor
structure of the latter seems to complicate the analysis
considerably. For example, the structure of the fermionic thermal
self-energy is such that it leads to the appearance of collective
plasma modes, see \Cref{s:holes}.  Fortunately, we do not have to delve
into these complications for the derivation of the VIA source. In
fact, as we will see, we can expand both the self-energies and Green's
functions in helicity eigenstates. The spinor structure then drops out
of the KB equations, and the derivation and cancellation of the VIA
source is then in almost one-to-one correspondence to the bosonic case.

With a slight abuse of terminology, we will now refer to the two chiralities as different `flavors', because the chiralities have different interactions with the thermal plasma in a chiral
theory such as the SM. Once thermal corrections are included, the propagating
and interaction states are not aligned, which is essential for the 
VIA source.

\subsection{Lagrangian}

For definiteness, consider two Weyl fermions with left and right chirality coupled via a
background-dependent Dirac mass $M_0=M_0(v)$
\begin{align}
  {\cal L} &= i \bar \psi \slashed{\partial} \psi
  - \bar\psi M_0 \psi = i \bar \psi \slashed{\partial} \psi
  - \bar\psi_L m_{LR} \psi_R - \bar \psi_R m_{RL}\psi_L,
\end{align}
with $m_{LR} = (m_{RL})^*$, $P_R M_0 P_R= m_{LR}$ and
$P_L M_0 P_L= m_{RL}$. 
The vev insertion approximation consists of treating
$m_{LR} = (m_{RL})^*$ as a perturbation, and expanding the
KB-equations in this small quantity.

\subsection{KB equations}

The KB equations \cref{KBabF} for the Wightman functions $G^\lambda$ and time-ordered propagator $G^t$ are 
\begin{align}
\{\slashed{k},G^\lambda\}&= \e^{-i\diamond}\(
\{M , G^\lambda\} +\{\Pi^\lambda , G^h\} + \frac12([\Pi^> ,
                                          G^<] -[\Pi^< , G^>] ) \),
    \nn
  \\
 \frac12\{\slashed{k},G^t\}&= 1+\frac12\e^{-i\diamond}\( 
                                     \{M+ \Pi^t-\Pi^h , G^t\}-\Pi^<
                                    G^> -G^< \Pi^>\),
\end{align}
where we defined $M = M_0 + \Pi^h$ with $M_0$  the
tree-level Dirac mass matrix, and $\Pi^h$ the thermal mass.
The source
terms can be derived from the kinetic equation for the Wightman functions
\begin{align}
  \frac{i}2\{\slashed{\partial}_x, G^\lambda\}&  = \e^{-i\diamond}\(
[M, G^\lambda]+ [\Pi^\lambda,
                                           G^h]+
                                          \frac12\(\{\Pi^> , G^<\}-\{\Pi^< ,
G^>\}\) \) .
\end{align}
Integrating the left-hand side over 4-momentum gives
\cite{Riotto:1997vy} 
\begin{align}
\Tr  \left[ \int \frac{\dd^4 k}{(2\pi)^4} i\{\slashed{\partial}_x, G^\lambda(k,x)\} \right]&=
\lim_{r\to 0} \Tr\, \left[i (\slashed{\partial}_x
G^\lambda(r,x) +G^\lambda(r,x) \overleftarrow{\slashed{\partial}}_x) \right] \nn \\
& =\partial_{x^\mu} \langle \bar \psi(x) \gamma^\mu \psi(x) \rangle = \partial_\mu
\langle J^\mu \rangle,
\end{align}
with $\Tr$ the trace over Dirac spinor indices.
The position-space limit $r \to 0$ corresponds to taking the
leading order derivative expansion in Wigner space, i.e. setting
$\e^{-i\diamond} =1$, which is picked out by the integration over
4-momenta.  The evolution of the left-handed current is then defined
as
\begin{align}
 \partial_\mu
\langle J_L^\mu \rangle \equiv \partial_{\mu} \langle \bar \psi_L(x)
  \gamma^\mu \psi_L(x) \rangle  = \Tr \Big[\int \frac{\dd^4
  k}{(2\pi)^4} i\Big( & P_R \slashed{\partial}_xP_L
  G^\lambda(k,x)P_R \nn\\ & \qquad+P_L
  G^\lambda(k,x)P_R \slashed{\partial}_x P_L\Big) \Big].
\end{align}
Hence, the source term for the left-handed fields can be written as
\begin{align}
S_{LL} &= \Tr \Big[ \int \frac{\dd^4 k}{(2\pi)^4} \(
[M, G^>+G^<]+ [\Pi^>+\Pi^<,
                                           G^h]+
                                         \(\{\Pi^> , G^<\}-\{\Pi^< ,
  G^>\}\) \)_{LL} \Big]\nn \\
  &\equiv \int \frac{\dd^4 k}{(2\pi)^4} \bar S_{LL},
  \label{SL_fermion}
  \end{align}
  with $(A B)_{LL} = \sum_{I=L,R} A_{LI}B_{IL}$. Note in this respect
  that for the propagator $G_{LL} = P_L G P_R$ while for the
  mass/self-energy $M_{LL} = P_R M P_L, \, \Pi_{LL} = P_R \Pi P_L$. In
  the last line we defined the integrand $\bar S_{LL}$ for future
  reference.
  
\subsection{Thermal corrections}

Let's first consider massless fermions and set the Dirac mass to
zero. This would be the starting point in the vev-insertion
approximation, where at leading order the off-diagonal vev-dependent
mass terms are set to zero. The self-energies are flavor-diagonal,
 and we approximate them with the thermal self-energies
that satisfy the KMS relations \cref{KMS}.  We can then asses the
effects of the self-energies in the diagonal basis of retarded and
advanced propagators using \cref{V}, where we work at leading order in
the derivative expansion.

The self-energies for the left and right chiralities may differ in a
chiral theory.  For a massless fermion
the hermitian and anti-hermitian self-energy can be written as
\begin{align}
\Pi^h &=P_R\( a_L \slashed{k}+b_L \slashed{u} \)P_L +P_L\( a_R
        \slashed{k}+b_R
        \slashed{u} \)P_R,
\nn \\
\Pi^{\cal A} &=P_R\(2 \bar \Gamma_L \gamma_0 \)P_L +P_L \(2 \bar
               \Gamma_R \gamma_0 \)P_R,
               \label{Pi_F}
\end{align}
with $a_I, b_I$ Lorentz invariants and functions of $k =|\vec k|$. In vacuum $b_I=0$ by Lorentz invariance, but at finite temperature the
center of mass frame of the plasma introduces a special Lorentz
frame. In the plasma rest frame the plasma four velocity is
$u^\mu =(1,0,0,0)$ and $\slashed{u} = \gamma^0$. 
The anti-hermitian
self-energy is calculated in the zero-momentum limit
\cite{Braaten:1992gd}, and can then be related to the decay
rate. Away from this limit, the expression is likely of the same
general form as $\Pi^h$, but for simplicity we will stick to the
simpler expression for $\Pi^{\cal A}$ in \cref{Pi_F}.

It will be useful to work in the basis of 2-dimensional Weyl spinors, and we define
\begin{align}
 M_0 = \( \begin{array}{cc}
                     m_{RL} &0 \\
                     0 &m_{LR} 
                                 \end{array} \)   , \quad
  \Pi ^{ab} = \( \begin{array}{cc}
                     0&\Pi ^{ab}_{RR}\\
                     \Pi ^{ab}_{LL} & 0
                   \end{array} \),\quad
  G^{ab} = \( \begin{array}{cc}
                     G_{LR}^{ab}&G^{ab}_{LL}\\
                   G^{ab}_{RR}&  G_{RL}^{ab}
                   \end{array} \),
\end{align}                                   
with the full matrices $M_0,\Pi,G$ all 4-by-4 in Dirac spinor space,
and the sub-blocks $\Pi_{IJ},G_{IJ},M_{IJ}$ 2-by-2 matrices in Weyl
spinor space.  Note that this is a change of notation, as below
\cref{SL_fermion} the same notation was used for the projection of
propagators and self-energies in Dirac space.   From now on, all
propagators, self-energies and masses will be given in Weyl space, and
we do not expect this to lead to confusion. 

We introduce the helicity projection operators 
\be
P^\pm = \frac12(1\pm \vecS \sigma \cdot \hat k),
\ee 
which satisfy
$P^+ P^-=0$, $(P^\pm)^2 =P^\pm$ and $P^+ + P^-=1$.  The hermitian
self-energy  can then be written as follows
\be
\Pi^h_{LL} =\Pi^h_{LL,+} P^+ + \Pi^h_{LL,-} P^-, \quad
\Pi^h_{RR} =\Pi^h_{RR,-} P^+ +\Pi^h_{RR,+} P^-,
\ee
with $\Pi^h_{II,\pm} = a_I K_\pm +b_I$ and $K_\pm = k^0 \pm k$. To
leading order in the derivative expansion the equation for the
retarded/advanced Green's function is
$ \frac12 \{ \slashed{k} ,G^\alpha \}- \frac12 \{\Pi^\alpha,G^\alpha\}
=1 $.
We rewrite $\slashed{k}$ in the Weyl basis
\begin{align}
  \slashed{k} &= \( \begin{array}{cc}
                     0& k^0 - \vec k \cdot \vecS \sigma \\
                     k^0 + \vec k \cdot \vecS \sigma & 0
                   \end{array} \)=\( \begin{array}{cc}
                     0& K_-P^+ + K_+ P^-  \\
                 K_+ P^+ +K_-P^- & 0
                                     \end{array} \).
                                   \label{slash_k}
\end{align}
The solution for the inverse retarded propagator becomes
\begin{align}
  (G_{LL}^r)^{-1}&=( K_- - \Pi^h_{LL,-} +2i \bar \Gamma_{L} )P^- +(
K_+  - \Pi^h_{LL,+} +2i \bar \Gamma_{L} )P^+, \nn \\
(G_{RR}^r)^{-1}&=( K_+ - \Pi^h_{RR,+} +2i \bar \Gamma_{R} )P^- +(
K_-  - \Pi^h_{RR,-} +2i \bar \Gamma_{R} )P^+.
\end{align}
Inverting this relation  gives
\be
G_{LL}^r = G_{LL,+}^r P^+  + G_{LL,-}^r P^-, \quad
G_{RR}^r = G_{RR,-}^r P^+  + G_{RR,+}^r P^-, 
\ee
with
\be
G^r_{II,\pm} = (K_\pm -\Pi^h_{II,\pm} +2 i\bar \Gamma_I)^{-1} \equiv D_{I,\pm}^{-1}.
\ee
The advanced propagator is obtained by taking
$i\bar \Gamma_I \to - i\bar \Gamma_I$.

To re-express the results in the basis of Wightman functions, we first
derive the spectral function. Define
\begin{align}
\rho_{I,\pm} =\( G_{II,\pm}^r-G_{II,\pm}^a\)=  \frac{1}{D_{I,\pm}}- \frac{1}{D_{I,\pm}^*} = \frac{\bar\gamma_I}{D_{I,\pm} D_{I,\pm}^*}=\frac{
  \bar\gamma_I}
  {(K_\pm-\Pi^h_{II,\pm})^2 -\bar \gamma_I^2/4},
  \label{rho_F}
\end{align}
with $\bar\gamma_I = -4i\bar \Gamma_I$, where we use the barred
notation to avoid possible confusion with a Dirac matrix. Then
 $\rho_L
= \rho_{L,+} P^++\rho_{L,-} P^-$  and  $\rho_R
= \rho_{R,-} P^++\rho_{R,+} P^-$.
With this notation we write the self-energies as
\be \Pi_{II}^\lambda
=\bar \gamma_I g^\lambda_I,
\label{gammaF}
\ee with $g^\lambda$ defined in \cref{Grho}.  This is analogous to
  our choice for the bosonic case \cref{gamma}, and follows the
  assumptions in the VIA literature.  It is at this point that our
  treatment diverges from the analysis in
  Ref.~\cite{Kainulainen:2021oqs}, which is based on vector-like
  interactions with the bath, whereas \cref{gammaF} is expected in a
  chiral theory. The discussion on flavor dynamics in
  \cref{s:disc_flav} carries over to the fermionic case.

The diagonal Green's functions are expanded in projection operators 
\be
G^{ab}_{LL} = G^{ab}_{LL,+} P^++G^{ab}_{LL,-} P^-, \quad
G^{ab}_{RR} = G^{ab}_{RR,-} P^++G^{ab}_{RR,+} P^-.
\label{WightmanP1}
\ee
The Wightman functions are then
\be
G^\lambda_{II,\pm}= g^\lambda_{II}\rho_I^\pm.
\label{WightmanP2}
\ee
The (anti) time-ordered propagators $G_{II}^t = -(G^r-G^a)_{II}n_I
+G^r_{II}$ and $G_{II}^{\bar t} = -(G^r-G^a)_{II}n_I
-G^a_{II}$ can be written as
\begin{align}
  G^t_{II,\pm} 
              = \(- n_I + \frac{D_{I,\pm}^*}{\bar \gamma_I} \)
  \rho_I^{\pm}, \qquad                               
  G^{\bar t}_{II,\pm}  
                     = \(- n_I - \frac{D_{I,\pm}}{\bar \gamma_I} \) \rho_I^{\pm},
\end{align}
where we used \cref{rho_F}. 

The constraint equations at leading order in both the derivative
expansion and the VIA expansion can be solved to directly find the
Wightman functions and time-ordered propagators. This is what we have
shown explicitly for bosons in \cref{s:constraint_B}. We will
not repeat the exercise here for fermions, but it is straightforward
to check that indeed the solutions given above solve the leading order
constraint equations.

To calculate the source term and show that also in the fermionic case
there is a cancellation, we can work with the spectral function given in
\cref{rho_F}, which is a solution of the constraint equation at
leading order in VIA.  Although there is no need to determine the pole
structure and identify the particle and collective hole modes, this is
further discussed in \cref{s:holes}, to show that this is indeed the
same spectral function as used in the usual VIA approach.

\subsection{VIA approximation}

The propagators found in the previous subsection 
are the leading order propagators in the VIA expansion. The KB
equations have a similar form as those for bosons, and it is no
surprise that the (leading order) Green's functions found in the
previous subsection have the same structure as those for
bosons as well. Likewise, as we will show, the cancellation
of the source term derived from the kinetic equation goes along the
same lines as what we have already seen for bosons.  In this
subsection  we will discuss the VIA approximation, and in the next
subsection we derive the source from the resummed propagators.

The propagators at 1st and 2nd order in VIA can be expressed in terms
of these zeroth order propagators by means of the Schwinger-Dyson \cref{G12}, which becomes  in Weyl space
\begin{align}
  G^{ab}_{(1),LR}&= m_{LR}\, \sum_c c\( G^{ac}_{LL,+} 
                  G^{cb}_{RR,-} P_+ + G^{ac}_{LL,-} 
                  G^{cb}_{RR,+} P_-\),\nn \\
 G^{ab}_{(2),LL} &= |m_{LR}|^2 \sum_{cd} cd \( G^{ac}_{LL,+} 
                    G^{cd}_{RR,-} G^{db}_{LL,+} P_++G^{ac}_{LL,-} 
                    G^{cd}_{RR,+} G^{db}_{LL,-} P_- \).
\end{align}
In this subsection, all propagators without a $(..)$ subscript are again to
be understood as the zeroth order propagator.

We can then evaluate the source term in the VIA expansion. Let's focus
on the left-handed source \cref{SL_fermion} at 2nd order in VIA.  We
insert $1 = (P^-+P^+)$ in the trace, and denote the projections of the
source onto $P^\pm$ by $\bar S^\pm$. We concentrate here on
$\bar S_{LL}^{+}$, but it is straightforward to check that all other diagonal
source contributions $S_{II}^\pm$ vanish as well.  At 2nd order in
VIA, 
\begin{align}
\bar S_{LL}^{(2)+} &= \Tr \left[ P^+ \(
[M_0, G_{(1)}^>+G_{(1)}^<]+
                                         \(\{\Pi^> , G_{(2)}^<\}-\{\Pi^< ,
  G_{(2)}^>\}\) \)_{LL} \right],
\end{align}
with now $\Tr$ denoting the trace in Weyl space, and
$(A B)_{LL} = \sum_{I=L,R} A_{LI}B_{IL}$ in Weyl space
notation. 
$M_0$
is the tree-level (Dirac) mass.

Consider first the commutator term
\begin{align}
 \bar S^{(2)+}_{M,LL} & \equiv \Tr \left [P^+ 
[M_0, G_{(1)}^\lambda] \right]=\Tr\left[ m_{LR} P^+ (G^>_{(1),RL}+G^<_{(1),RL})-
                P^+  (G^>_{(1),LR}+G^<_{(1),LR}) m_{RL} \right]\nn \\
  & = \Tr [P^+] |m_{LR}|^2\[  (G^t+G^{\bar t})_{LL,+} (G^> -G^<)_{RR,-}-
    (G^t+G^{\bar t})_{RR,-} (G^> -G^<)_{LL,+}\]  \nn \\
  &=
  - 2 |m_{LR}|^2 \rho_L^+\rho_R^- (n_L-n_R),
\end{align}  
 where we used that
$(G^t_\pm +G^{\bar t}_\pm)_{II}= (1-2n_I) \rho_I^\pm$ and
$(G^> -G^<)_{\pm,II} = \rho_I^\pm$. Further $\Tr (P^+)=1$. This has
exactly the same structure as in the bosonic case \cref{S2M_B}, except
for an overal sign difference because of the difference between bose
and fermi statistics.

The collision term contributes
\begin{align}
\bar S^{(2)+}_{C,LL} &\equiv
             2\, \Tr \left[P^+ \(G_{(2),LL}^< \Pi_{LL}^> -G_{(2),LL}^> \Pi_{LL}^<\)\right]
             =2 | m_{LR}|^2\, ( {\rm term}_1+ {\rm term}_2), \label{eq:fermcol}
\end{align}
where we have used the KMS relations to already cancel some
contributions, fully analogous to what happens in the bosonic
expression \cref{S2M_C}.  The two remaining terms are then
\begin{align}
  {\rm term}_1& \equiv
\Tr \[ P^+\( \Pi^<_{LL} G^>_{LL} G^<_{RR} G^>_{LL} -\Pi^>_{LL}
    G^<_{LL} G^>_{RR} G^<_{LL}\) \]
                \nn \\
  &=\Tr[ P^+] (\Pi^<_{LL} G^>_{+,LL} G^<_{-,RR} G^>_{+,LL} -\Pi^>_{LL}
         G^<_{+,LL} G^>_{-,RR} G^<_{+,LL})  
         \nn \\ &= - \bar \gamma_L  (\rho_L^+)^2 \rho_R^- n_L (1-n_L)(n_L-n_R),
\end{align}
and
\begin{align}
{\rm term}_2 &\equiv \Tr \[ P^+  (\Pi^<_{LL}G^>_{RR} -\Pi^>_{LL}G^<_{RR}  )
          G^{\bar t}_{LL}   G^t_{LL}\] 
  = (\Pi^<_{LL}G^>_{-,RR} -\Pi^>_{LL}G^<_{-,RR}  )
          G^{\bar t}_{+,LL}   G^t_{+,LL} 
    \nn \\ & =- \bar \gamma_L  (\rho_L^+)^2 \rho_R^-\(n_L -
             \frac{D^*_{L,+}}{\bar \gamma_L} \) \(n_L +
             \frac{D_{L,+}}{\bar \gamma_L} \) \( n_L -n_R\) .
\end{align}
Just as for bosons, the result simplifies when we add the two terms
\begin{align}
({\rm term}_1+{\rm term}_1)& = 
                  \rho_L^+ \rho_R^-(n_L-n_R) 
\end{align}
where we used that $D_{L,+}-D^*_{L,+}+\bar \gamma_L=0$.
The final result is then 
\begin{align}
 \bar  S^{(2)+} &=\bar S^{(2)+}_{M,LL} +\bar S^{(2)+}_{C,LL} = 
                         -2 |m_{LR}|^2 \rho_L^+\rho_R^- (n_L-n_R) \nn
  +
                       2 | m_{LR}|^2  \rho_L^+\rho_R^-  (n_L-n_R) \nn \\
  &=0.
\end{align}                 
We can likewise calculate the source term projected onto negative
helicity using the $P^-$ operator. This amounts to interchanging the
plus and minus labels, and thus also gives a vanishing source term.
We conclude that the 2nd order VIA source cancels also for
fermions. In the next subsection we will show that this cancellation
occurs at all orders in VIA.

We can again compare with the usual derivation of the fermionic source
term in the literature \cite{Riotto:1995hh,Lee:2004we}, which is
derived expanding the self-energy to 2nd order in VIA. In this
approach only the collision term contributes, and the contribution is
equal to our result for the collision term. To write that contribution
to the source in a more familiar form, perform an inverse Wigner
transform on \cref{eq:fermcol} and expand the masses in the integrand
to linear order, just as we did in \cref{s:kinetic_B} for the bosonic
case.  However, as noted before, this standard non-IPI approach misses
the cancellation of the source term.

\subsection{Resummed Green's functions and source term}

We will now derive the fully resummed source term.
 The KB constraint equations at leading derivative order are
\begin{align}
\{\slashed{k},G^\lambda\}&= 
\{M , G^\lambda\} +\{\Pi^\lambda , G^h\} + \frac12([\Pi^> ,
                                          G^<] -[\Pi^< , G^>] ), 
    \nn
  \\
 \frac12\{\slashed{k},G^t\}&= 1+\frac12\( 
                                     \{M+ \Pi^t-\Pi^h , G^t\} -\Pi^<
                                    G^> -G^< \Pi^>\).
\end{align}
We express all propagators and self-energies in the Weyl
basis as before, see \Cref{s:weyl} for explicit expressions.

We then project all four components in Weyl space onto the $P^+$ and
$P^-$ propagators, which gives two decoupled systems of equations. 
The solutions for
$G_{LL,+}$, $G_{LR,+}$ and $G_{RL,-}$ are
\begin{align}
  G_{LL,+}^\lambda &= \frac{1}{\D^+ \D^-}
                     \frac{\bar\gamma_R \bar\gamma_L}{\rho^-_R}\(g_L^\lambda
                     +g_R^\lambda \frac{\rho^-_R}{\bar\gamma_L} |m_{LR}|^2 \),\nn\\
 G_{LR,+}^\lambda &= \frac{m_{LR}}{\D^+ \D^- } \(
\bar \gamma_R g_R^\lambda(K_+-\Pi^h_{LL,+}) +\bar\gamma_L g_L^\lambda (K_- -\Pi^h_{RR,-})
                   +\frac12 \bar\gamma_L \bar\gamma_R (g^\lambda_R-g^\lambda_L)\), \nn \\
   G_{RL,-}^\lambda &= \frac{m_{RL}}{\D^+ \D^- } \(
\bar \gamma_R g_R^\lambda(K_+-\Pi^h_{LL,+}) +\bar\gamma_L g_L^\lambda (K_- -\Pi^h_{RR,-})
                   +\frac12 \bar\gamma_L \bar\gamma_R (g^\lambda_L-g^\lambda_R)\),
\end{align}
with $\rho^\pm_I$ the
zeroth order spectral functions \cref{rho_F}, and
\begin{align}
&\D^\pm = (K_+ -\Pi^h_{LL,+} \pm\bar\gamma_L/2)(K_- -\Pi^h_{RR,-}
                \pm\bar\gamma_R/2)-|m_{LR}|^2.
                \label{Dplusmin}
\end{align}
In the limit of vanishing Dirac
masses \be \lim_{m_{LR},m_{RL} \to 0} \left\{ \frac{1}{D^+ D^-}
  \frac{\bar\gamma_R\bar\gamma_L}{\rho^-_R},\, \frac{1}{\D^+ \D^-}
  \frac{\bar\gamma_R\bar\gamma_L}{\rho^+_L}
\right\}=\{\rho_L^+,\,\rho_R^-\}, \ee
and the 0th order Wightman functions are retrieved.  The fermionic
solutions have a similar structure as their bosonic counterpart
\cref{Wresummed_B}. Pinch singularities in the separate
mass commutator and collision term are again avoided, as the
denominators in the propagator are shifted by the non-zero Dirac mass
\cref{Dplusmin}.

Now we calculate the left-handed source \cref{SL_fermion} projected onto $P^+$:
\begin{align}
\bar S^+_{LL} &= \frac 1 2 \Tr P^+\(
[M, G^>] +[M, G^<]+ [\Pi^>,
                                           G^h]+ [\Pi^<,G^h]+
                                          \(\{\Pi^> , G^<\}-\{\Pi^< ,
             G^>\}\) \)_{LL} \nn \\
  & = \frac 1 2\Tr(P^+) \( m_{LR} (G_{RL,-}^>+G^<_{RL,-})- (G_{LR,+}^>+G^<_{LR,+})
    m_{RL}
    + 2(\Pi^>_{LL} G^<_{LL,+} -\Pi^<_{LL,+}G^>_{LL} ) \)\nn \\
    &=0. \label{eq:SfermLL}
  \end{align}
Only the commutator with the tree-level Dirac mass $M_0$ 
and the collision term contribute, and they add up to zero. There is thus
no fermionic VIA source at all orders in the vev-insertion expansion.


\section{Conclusion}\label{s:conclusion}

In this work we have shown that the resonantly enhanced source term
traditionally derived in the vev insertion approximation gets
cancelled exactly when all terms in the Kadanoff-Baym equations at
leading order in the derivative expansion are included. The
  derivation of the original VIA source relied crucially on the
  existence of non-trivial flavor dynamics, i.e. a non-alignment
  between the propagating and flavor eigenstates. In the case of
  fermions, the term `flavor' is (ab)used to refer to the two
  different chiralities, which obtain distinct thermal corrections
  from the plasma. In our analysis we have followed the original
  VIA literature, and allowed for flavor-dependent interactions (chiral
  interactions for fermions) with the thermal bath. This generalizes
  the results of \cite{Kainulainen:2021oqs}, which considered vector
 -like interactions.

We suspect that the
cancellation of the contribitions to the source went unnoticed in the original VIA literature, because of
the use of 1-particle-reducible (non-1PI) self-energies. Instead, our
analysis relies solely on 1PI self-energies, and uses a systematic
perturbative expansion. In hindsight our result may have been foreseen.
Indeed, the VIA expansion source arises at leading order in the
derivative expansion, at which order no distinction between a constant
and an evolving background is made -- and certainly no source term is
expected in the former case.

Interestingly, the problem with the original derivation of the VIA
source term does {\it not} stem from a pathology in the vev insertion
expansion itself, or from the existence of so-called pinch
singularities (which are absent when the fully resummed propagator
solutions are used), but instead from the approach based on non-IP1 diagrams
that obscures a proper counting of all contributions.

To show the cancellation of the source one must include the
collision term in the kinetic equation, which is non-zero because of
the non-trivial flavor dynamics.  The assumption is that the
interactions with the thermal bath are fast, and to a good
approximation the bath can be regarded as a collection of particles in
flavor eigenstates. The thermal self-energies then satisfy
KMS relations \cref{KMS}, formulated in terms of the number densities of flavor
eigenstates. The spectral function, on the other hand, contains
information on the mass spectrum of the system. If the mass and flavor
states are not aligned, the Wightman functions no longer satisfy KMS
relations in terms of the number densities of flavor eigenstates, and
thus the collision term does not necessarily vanish.  Our explicit
expressions for the collision term, see e.g. \cref{B_collision},
indeed vanish in the limit that the flavor off-diagonal mass term
$m_{LR}$ is set to zero, and flavor mixing is absent.

Computations of the baryon asymmetry using the VIA source suggested a
rosier picture of the feasibility of EWB in a number of BSM models.
The larger yield was mainly due to the CP-violating VIA source containing
only one spacetime derivative, and consequently there is no partial cancellation in
particle asymmetries when integrated
over the bubble wall, in sharp contrast to the semi-classical and
flavor oscillation sources \cite{Cline:2021dkf}.  We do expect that
the VIA approximation gives a non-zero source term at the next (first)
order in the derivative expansion, and that the inclusion of thermal
effects may lead to resonant enhancement. But it remains to be seen
whether this will be enough to save large classes of models from the
ever tightening constraints imposed by the non-observation of EDMs.

\section*{Acknowledgements}
MP is supported by the Netherlands Organization for Scientific
Research (NWO). JvdV is supported by the Deutsche Forschungsgemeinschaft under Germany's Excellence Strategy -- EXC 2121 ``Quantum Universe'' -- 390833306. GW is supported by World Premier International Research Center Initiative (WPI Initiative), MEXT, Japan. We thank Jordy de Vries and Thomas Konstandin for helpful discussions. This work benefitted from discussions at the 2021 Lorentz Center workshop `Computations that Matter'.

\newpage
\appendix

\section{Weyl basis}
\label{s:weyl}

In this appendix we list all propagators and self-energies expressed
in the Weyl basis, defined as 2-by-2 subblocks of the matrices in
Dirac space.
\begin{align}
  \slashed{k} &= \( \begin{array}{cc}
                     0& k^0 - \vec k \cdot \vecS \sigma \\
                     k^0 + \vec k \cdot \vecS \sigma & 0
                   \end{array} \)=\( \begin{array}{cc}
                     0& K_-P^+ + K_+ P^-  \\
                 K_+ P^+ +K_-P^- & 0
                                     \end{array} \),   \nn \\
   \Pi^h &= \( \begin{array}{cc}
                     0&  \! a_R (k^0 \! - \! \vec k \cdot \vecS \sigma) \! +\! b_R \! \\
                 \!   a_L (k^0 \!+\!  \vec k \cdot \vecS \sigma) \! \!+b_L \! & 0
                   \end{array} \) \!=\!\( \begin{array}{cc}
                     0& \!\Pi^h_{RR,-}P^+ \!+ \!\Pi^h_{RR,+} P^- \! \\
             \!   \Pi^h_{LL,+} P^+\! +\!\Pi^h_{LL,-}P^-\! & 0
                   \end{array} \),   \nn \\
               M_0 &= \( \begin{array}{cc}
                     m_{RL} &0 \\
                     0 &m_{LR} 
                                 \end{array} \) ,          \nn \\
  \Pi^\lambda& = \( \begin{array}{cc}
                     0&\Pi^\lambda_{RR}\\
                     \Pi^\lambda_{LL} & 0
                   \end{array} \) = \( \begin{array}{cc}
                     0&\bar \gamma_{R} g^\lambda_R\\
                   \bar  \gamma_{L} g^\lambda_L & 0
                                       \end{array} \),
    \nn \\
  G^{ab}& = \( \begin{array}{cc}
                     G_{LR}^{ab}&G^{ab}_{LL}\\
                   G^{ab}_{RR}&  G_{RL}^{ab}
                   \end{array} \) = \( \begin{array}{cc}
                      G^{ab}_{LR,+}P^+ + G^{ab}_{LR-}P^-& \, G^{ab}_{LL,+}P^+ + G^{ab}_{LL-}P^-\\
                     G^{ab}_{RR,-} P^++ G^{ab}_{RR,+} P^- & \,  G^{ab}_{RL,-} P^++ G^{ab}_{RL,+} P^-
                                       \end{array} \),
                                                            \label{matrices}
\end{align}                                                   
where we defined $K_\pm = k^0 \pm k$. 

\section{Particle and hole modes}
\label{s:holes}

In this appendix we give the connection between our propagator
solutions to the usual form of the Dirac propagators -- with particle
and hole modes -- used in the derivation of the VIA source in the
literature, see in particular Ref.~\cite{Lee:2004we}. 
In contrast to the analysis in \Cref{s:fermions}, but in correspondence with \cite{Weldon:1999th,Lee:2004we}, we will work
with 4-dimensional Dirac spinors in this appendix.

The self-energies for the left and right chiralities may differ in a
chiral theory and we split
\be
\Pi^\alpha = \Pi_{LL}^\alpha +\Pi_{RR}^\alpha =P_R \Pi^\alpha P_L
+P_L \Pi^\alpha P_R,
\ee
with $\Pi^\alpha = \Pi^h \mp i\Pi^{\cal A} $.  For a massless fermion
the self-energy can be written as
\be
\Pi_{II}^h =P_J\( a_I \slashed{k}+b_I \slashed{u} \)P_I
,
\qquad
\Pi_{II}^{\cal A} =P_J \(2 \bar \Gamma_I \gamma_0 \)P_I,
\ee
with $J \neq I$.
The solution to the leading order constraint equation for the
retarded/advanced Green's function,  $(G^r)^{-1}= (G_{LL}^r)^{-1}+ (G_{RR}^r)^{-1}
=  P_R (G^r)^{-1}P_L+P_L  (G^r)^{-1}P_R$, is
\begin{align}
(G_{II}^r)^{-1} &= P_J\( (A_{I,0} +  2i \bar \Gamma_I)\gamma^0 -A_{I,s} \vecS \gamma \cdot
\hat{\vec k}\)P_I= D_{I,+} \Lambda^{JI}_+ +D_{I,-} \Lambda^{JI}_-  ,
\end{align}
with $I=L,R$. In the middle expression the coefficients are
\be
A_{I,0} = k_0(1-a_I) -b_I ,\quad
A_{I,s} = k(1-a_I),
\ee
and on the RHS we rewrote this defining
\begin{align}
  D_{I,\pm} = A_{I,0} +2i \bar \Gamma_I \mp A_{I,s}, \qquad
  \Lambda_\pm = \frac12(\gamma^0 \pm
\vecS\gamma \cdot \hat{\vec k}), \qquad \Lambda_\pm^{IJ}= P_I
  \Lambda_\pm P_J \;\; {\rm for}\;\; I \neq J.
\end{align}
The advanced propagator is of the same form with $i\bar \Gamma \to - i
\bar \Gamma$.

To invert the propagator we use that $\Lambda^\pm \Lambda^\pm=0$ and
$\Lambda^+\Lambda^- +\Lambda^-\Lambda^+=1$. Then
$G^r = G_{LL}^r+ G_{RR}^r=P_L G^r P_R + P_R G^r P_L$ with
\begin{align}
G_{II}^r   =\frac{\Lambda^{IJ}_{-}}{D_{I,+}} +
  \frac{\Lambda^{IJ}_+}{D_{I,-}}.
  \label{Gr_ferm}
\end{align}
The poles of the
propagators are located at $k_0 =\E_{I,p}$ (and $k_0 =-\E^*_{I,p}$),
which is the solution of $D_{I,+}=0$, and at $k_0 = \E_{I,h}$ (and
$k_0 =-\E^*_{I,h}$), which is the solution of $D_{I,-}=0$.  
The real
parts of the poles $E_{I} = {\rm Re} (\E_{I})$ correspond to the
particle and hole energies (antiholes and antiparticles) respectively,
as indicated by the subscript. 
For the retarded propagator the poles
of the propagator have negative imaginary parts, and in the narrow
width approximation
\be
\E_{I,p} = E_{I,p} - i \Gamma_{I,p}, \quad \E_{I,h} = E_{I,h} -i \Gamma_{I,h},
\label{Epi}
\ee
with $\Gamma_I = Z_{I,p}\bar \Gamma_I$ and
\be
Z_{p,h} =(\partial_{k_0} D_\pm)^{-1}|_{k_0 =\E_{p,h}}= \frac{E^2_{p,h}
-k^2}{2m_T^2},
\ee
the residues of the poles \cite{Weldon:1999th, Lee:2004we}. The
retarded propagator can then be written as
\begin{align}
G_{II}^r = \Big( \frac{Z_{I,p}}{k_0 -\E_{I,p}} &+ \frac{Z_{I,h}^*}{k_0 +\E_{I,h}^*}  -f_I(k_0,k)\Big)
\Lambda_-^{IJ} \nn \\ &+\Big( \frac{Z_{I,h}}{k_0 -\E_{I,h}} + \frac{Z_{I,p}^*}{k_0 +\E_{I,p}^*}  +f_I^*(-k_0^*,k)\Big)
\Lambda_+^{IJ}.
\end{align}
The function $f_I$ gives the non-pole part of the propagator, which
is subdominant near a resonance.
The poles at $k_0
= -\E^*_{I,p},-\E^*_{I,h}$ correspond to excitations of anti-particles and
anti-holes.  The spectral function is  $\rho =\( G^r-G^a\)
= \rho_{L}+\rho_{R}= P_L\rho P_R+ P_R \rho P_L$ with,
\be
\rho_I =P_I\( G^r-G^a\) P_J = \Lambda_-^{IJ}\rho_{I,+}+\Lambda_+^{IJ}
\rho_{I,-},
\ee
and
\begin{align}
\rho_{I,+} =  \frac{Z_{I,p}}{k_0-\E_{I,p}}-\frac{Z_{I,p}^*}{k_0-\E_{I,p}^*}+\frac{Z_{I,h}^*}{k_0+\E_{I,h}^*}-\frac{Z_{I,h}}{k_0+\E_{I,h}}
  +...
\end{align}
and $\rho_{I,-} (k_0,k) = (\rho_{I,+} (-k_0^*,k) )^*$, and the elipses
denoting the sub-dominant non-pole contributions (see
\cite{Bellac:2011kqa} for their explicit form).  The Wightman
functions, generalized for non-zero chemical potential in
Ref.~\cite{Lee:2004we} \footnote{Note that Ref.~\cite{Lee:2004we} uses
  $\bar m_T^2 = 2m_T^2$ and there is no factor
  $1/2$ in the residue.  Their dispersion relation (A.18) differs
   a factor 2 compared to \cite{Bellac:2011kqa,Weldon:1982bn}
  given their definition of the thermal mass $\bar m_T^2 = g^2 C_2(r)T^2/2$.
}, are
\begin{align}
  G_{II}^\lambda=g_I^\lambda\( \Lambda_-^{IJ} \rho_{I,+}+\Lambda_+^{IJ} \rho_{I,-}\),
\end{align}  
with $n_I(k^0-\mu_I)= (e^{(k_0 -\mu_I)/T}+1)^{-1}$ the
Fermi-Dirac distribution.

In the hard thermal loop approximation
\begin{align}
  A_{I,0} 
=k_0  -\frac{m_I^2}{k_0}x Q_0(x), \quad
A_{I,s} 
= k+\frac{m_I^2}{k}(1-xQ_0(x)),
\label{A_HTL}
\end{align}
with $Q_0 = \frac12 \ln \frac{x+1}{x-1}$ the Legendre function of the
2nd kind, $x = k_0/k$, and $m_I^2 \equiv g^2 C^I_2T^2/8$, with $C_2^I$ denoting the quadratic Casimir operator of the gauge group under which particle $I$ is charged.
In the limit
$k \to 0$ of zero momentum, $xQ_0 \to 1 $, which gives $E_{I,p} =
E_{I,h} = m_I$ 
and $Z_{I,p} =Z_{I,h} = 1/2$. This motivates the definition of the
thermal mass $m_I$. 

In the limit $k\to \infty$ the hole contribution becomes subdominant
$Z_{I,h} \ll Z_{I,p}$.  Already for $k \gtrsim m_I$ this becomes a
good approximation and this limit was often used in the literature on the
VIA source, to avoid the complication of hole modes.  The particle pole is located at
$E_{I,p} = k+m^2_I/k$, which is the high momentum limit of a particle
with energy $E_{I,p} = \sqrt{k^2+2m_I^2}$, and $Z_p =1$.  In this case
it is more useful to define the thermal mass as $\bar m_I^2 =2m_I^2$.

\newpage
\bibliographystyle{jhep} 
\bibliography{myrefs}

\end{document}